\def\be{\begin{equation}}
\def\ee{\end{equation}}
\def\ba{\begin{array}}
\def\ea{\end{array}}

\documentclass[groupedaddress,aps,superscriptaddress,showpacs,nofootinbib]{revtex4}%
\usepackage{epsfig,dsfont,amssymb,amsmath,amsthm,amsfonts,amsbsy,mathrsfs}
\usepackage{graphicx, color}
\usepackage{amsmath}
\usepackage{amssymb}
\usepackage{mathdots}
\usepackage{float}
\usepackage{cases}
\usepackage{graphpap}
\usepackage{wrapfig}
\usepackage{tikz}
\usepackage{indentfirst}
\usepackage{picinpar,graphicx}
\begin{document}
\title{Improved Uncertainty Relation in the Presence of Quantum Memory}
\author{Yunlong Xiao}%mathxiao123@gmail.com
%\thanks{equally contributed to this work }
\affiliation{School of Mathematics, South China University of Technology, Guangzhou, Guangdong 510640, China}
\affiliation{Max Planck Institute for Mathematics in the Sciences, 04103 Leipzig, Germany}
\author{Naihuan Jing\footnote{Corresponding Author: jing@ncsu.edu}}
\affiliation{Department of Mathematics, Shanghai University, Shanghai 200436, China}
\affiliation{Department of Mathematics, North Carolina State University, Raleigh, NC 27695, USA}
\author{Shao-Ming Fei}%feishm@mail.cnu.edu.cn
\affiliation{School of Mathematical Sciences, Capital Normal University, Beijing 100048, China}
\affiliation{Max Planck Institute for Mathematics in the Sciences, 04103 Leipzig, Germany}
\author{Xianqing Li-Jost}%xli-jost@mis.mpg.de
\affiliation{Max Planck Institute for Mathematics in the Sciences, 04103 Leipzig, Germany}

%\title{The improved entropic uncertainty principle}

\begin{abstract}
Berta {\it et al}'s uncertainty principle in the presence of quantum memory [M. Berta {\it et al.}, Nat. Phys. \textbf{6}, 659 (2010)]
reveals uncertainties with quantum side information between the observables.
In the recent important work of Coles and Piani
[P. Coles and M. Piani, Phys. Rev. A. \textbf{89}, 022112 (2014)],
the entropic sum is controlled by the first and second maximum
overlaps between the two projective measurements. We generalize the entropic uncertainty relation in the presence of quantum memory
and find the exact dependence on all $d$ largest overlaps between two measurements on any $d$-dimensional Hilbert space.
Our bound is rigorously shown to be strictly tighter than previous entropic bounds in the presence of quantum memory, which have potential applications to quantum cryptography with entanglement witnesses and quantum key distributions.
\end{abstract}

 %{03.67.-a, 02.20.Hj, 03.65.-w}

\maketitle

Heisenberg's uncertainty principle \cite{Heisenberg} plays a central role in physics and marks
 a distinguished characteristic of quantum mechanics. The principle bounds the uncertainties of measurement outcomes of
 two observables, such as the position and momentum of a particle. This shows the underlying difference of quantum mechanics from classical
 mechanics where any properties of a physical object can be quantified exactly at the same time.
In Robertson's formulation \cite{Robertson}, the product of the standard deviations
(denoted by $\Delta(R)$ for the observable $R$) of the measurement of two observables $R$ and $S$ is controlled by their commutator:
\begin{align}\label{e:Robertson}
\Delta R\Delta S\geqslant\frac{1}{2}|\langle[R, S]\rangle|,
\end{align}
where $\langle\cdot|\cdot\rangle$ is the expectation value. The relation implies that it is impossible to simultaneously measure exactly a pair of incompatible (noncommutative) observables.

In the context of both classical and quantum information sciences, it is more natural to use entropy to quantify uncertainties \cite{Birula, Marco}.
The first entropic uncertainty relation for position and momentum was given in \cite{BBM} (which can be shown to be equivalent to
Heisenberg's original relation). Later Deutsch \cite{Deutsch}
found an entropic uncertainty relation for any pair of observables.
An improvement of Deutsch's entropic uncertainty relation was subsequently conjectured by Kraus \cite{Kraus} and later proved by Maassen and Uffink \cite{Maassen} (we use base $2$ $\log$ throughout this paper),
\begin{align}\label{e:MU}
H(R)+H(S)\geqslant\log \frac1{c_{1}},
\end{align}
where $R=\{|u_{j}\rangle\}$ and $S=\{|v_{k}\rangle\}$ are two orthonormal bases on $d$-dimensional Hilbert space $\mathcal{H}_{A}$, and $H(R)=-\sum_{j}p_{j}\log p_{j}$ is the Shannon entropy of the probability distribution $\{p_{j}=\langle u_{j}|\rho_{A}|u_{j}\rangle\}$ for state $\rho_{A}$ of $\mathcal{H}_{A}$ (similarly for $H(S)$ and $\{q_{k}=\langle v_{k}|\rho_{A}|v_{k}\rangle\}$). The number $c_{1}$ is the largest
overlap among all $c_{jk}=|\langle u_{j}|v_{k}\rangle|^{2}$ ($\leqslant 1$) between the projective measurements $R$ and $S$.

One of the important recent advances on uncertainty relations is to allow the measured quantum system to be correlated with its environment
in a non-classical way, for instance, picking up quantum correlations such as entanglement in quantum cryptography.
Historically, the entropic uncertainty relations have inspired initial formulation of quantum cryptography.
But the uncertainty relations
in the absence of quantum memory did not leave any chance
for an eavesdropper to have access to the quantum correlations. Therefore the ``classical'' uncertainty
relations without quantum side information cannot be utilized to improve cryptographic security directly.
Berta {\it et al.} \cite{B} bridged the gap between cryptographic scenarios and the uncertainty principle, and derived this landmark uncertainty relation for measurements $R$ and $S$ in the presence of quantum memory $B$:
\begin{align}\label{e:Berta}
H(R|B)+H(S|B)\geqslant\log \frac1{c_{1}}+H(A|B),
\end{align}
where $H(R|B)=H(\rho_{RB})-H(\rho_{B})$ is the conditional entropy with $\rho_{RB}=\sum_{j}(|u_{j}\rangle\langle u_{j}|\otimes I)(\rho_{AB})(|u_{j}\rangle\langle u_{j}|\otimes I)$ (similarly for $H(S|B)$),
and $d$ is the dimension of the subsystem $A$.
The term $H(A|B)=H(\rho_{AB})-H(\rho_{B})$ appearing on the right-hand side is related to the entanglement between the measured particle $A$ and the quantum memory $B$.

The bound of Berta {\it et al.} has recently been upgraded by Coles and Piani \cite{Coles}, who have shown a remarkable bound
in the presence of quantum memory
\begin{align}\label{e:Coles}
H(R|B)+H(S|B)\geqslant\log\frac1{c_{1}}+\frac{1-\sqrt{c_{1}}}{2}\log\frac{c_{1}}{c_{2}}+H(A|B),
\end{align}
where $c_{2}$ is the second largest overlap among all $c_{jk}$ (counting multiplicity) and other notations are the same as in Eq.(\ref{e:MU}).
As $1\geqslant c_1\geqslant c_2$, the second term in Eq.(\ref{e:Coles}) shows that
the uncertainties depend on more detailed information of the transition matrix or overlaps between the two bases. The Coles-Piani bound
offers a strictly tighter bound than the bound of Berta {\it et al.} as long as $1>c_1>c_2$.
The goal of this paper is to report a more general and tighter bound for the
entropic uncertainty relation with quantum side information.

As reported in \cite{Rudnicki}, there are some examples where the bounds such as $B_{Maj}$ and $B_{RPZ}$ based
on majorization approach outperform the degenerate form of Coles and Piani's new bound in the special case when quantum memory
is absent.
However, it is unknown if these bounds and their approaches can be extended to allow for quantum side information (cf. \cite{Marco}). Therefore Coles and Piani's remarkable bound Eq. (\ref{e:Coles}) is still the strongest
lower bound for the entropic sum in the presence of quantum memory {\it up to now}.
In this paper, we improve the bound of Coles and Piani in the most general
situation with quantum memory. Moreover, our new general bound is proven stronger by rigorous mathematical arguments.

To state our result, we first recall the majorization relation between two probability distributions $P=%(p_{j})
(p_1, \cdots, p_d)$, $Q=(q_1, \cdots, q_d)$. The partial order $P\prec Q$ means that $\sum_{j=1}^ip_j^{\downarrow}\leqslant \sum_{j=1}^iq_j^{\downarrow}$ for all $i=1, \cdots, d$. Here $^{\downarrow}$ denotes rearranging the components
of $p$ or $q$ in descending order. Any probability distribution vector $P$ is bounded by
$(\frac 1d, \cdots, \frac 1d)\prec P\prec (1, 0, \cdots, 0)=\{1\}$. For any two probability distributions $P=(p_j)$ and $Q=(q_k)$
corresponding to measurements $R$ and $S$ of the
state $\rho$, there is a state-independent bound of direct-sum majorization \cite{Rudnicki}: $P\oplus Q\prec\{1\}\oplus W$, where
$P\oplus Q=(p_1\cdots, p_d, q_1, \cdots, q_d)$ and $W=(s_{1}, s_{2}-s_{1}, \cdots, s_{d}-s_{d-1})$ is a special probability distribution
vector defined exclusively by the overlap matrix related to $R$ and $S$.
Let $U=(\langle u_{j}|v_{k}\rangle)_{jk}$ be the overlap matrix between the two bases given by $R$ and $S$,
and define the subset $\mathrm{Sub}(U, k)$ to be the collection of all size $r\times s$ submatrices $M$ such that
$r+s= k+1$. Following \cite{Rudnicki, Puchala} we define $s_{k}=\max\{\|M\|: M\in\mathrm{Sub}(U, k)\}$, where $\|M\|$
 is the maximal singular value of $M$.
Denote the sum of the largest $k$ terms in $\{1\}\oplus W$ as $\Omega_{k}=1+s_{k-1}$ and the $i$-th largest overlap among
$c_{jk}$'s as $c_{i}$. Meanwhile $s_0=0$, $s_1=\sqrt{c_1}$ and $s_d=1$. If follows from the basic definitions of $\Omega_{k}$ that the
following inequalities hold
\begin{align*}
1=\Omega_1\leqslant \Omega_2\leqslant\cdots\leqslant \Omega_{d+1}=\cdots=\Omega_{2d}=2,
\end{align*}
where we noted that $\Omega_2=1+\sqrt{c_1}$.

Our main result is the following entropic uncertainty relation that, much like Coles-Piani's bound,
accounts for the possible use of a quantum side information due to the entanglement between the measured
particle and quantum memory. For a bipartite quantum state $\rho_{AB}$ on
Hilbert space $\mathcal{H}_A\otimes\mathcal{H}_B$, we still use $H$ to denote the
von Neumann entropy, $H(\rho_{AB})=-\mathrm{Tr}(\rho_{AB}\log \rho_{AB})$.

\noindent\textbf{Theorem.} {\it Let $R=\{|u_{j}\rangle\}$ and $S=\{|v_{k}\rangle\}$ be arbitrary orthonormal bases of
the $d$-dimensional subsystem $A$ of a bipartite state $\rho_{AB}$. Then we have that
\begin{align}\label{e:result2}
H(R|B)+H(S|B)
\geqslant\log \frac1{c_{1}}+\frac{1-\sqrt{c_{1}}}{2}\log\frac{c_{1}}{c_{2}}+\frac{2-\Omega_{4}}{2}\log\frac{c_{2}}{c_{3}}
%\frac{2-\Omega_{6}}{2}\log\frac{c_{3}}{c_{4}}
+\cdots+\frac{2-\Omega_{2(d-1)}}{2}\log\frac{c_{d-1}}{c_{d}}+H(A|B),
\end{align}
where $H(R|B)=H(\rho_{RB})-H(\rho_{B})$ is the conditional entropy with $\rho_{RB}=\sum_{j}(|u_{j}\rangle\langle u_{j}|\otimes I)(\rho_{AB})(|u_{j}\rangle\langle u_{j}|\otimes I)$ (similarly for $H(S|B)$),
and
$H(A|B)=H(\rho_{AB})-H(\rho_{B})$.}

We remark that due to $\Omega_{d+1}=\cdots=\Omega_{2d}=2$,
the last (non-zero) term of formula (\ref{e:result1}) can be fine-tuned
according to the parity of $d$. If $d=2n$, it is $\frac{2-\Omega_{d}}{2}\log\frac{c_n}{c_{n+1}}$;
if $d=2n+1$, it is $\frac{2-\Omega_{d-1}}{2}\log\frac{c_{n}}{c_{n+1}}$.

\noindent\textbf{Proof.} For completeness we start from the derivation of the Coles-Piani inequality.
Observe that the quantum channel $\rho\rightarrow \rho_{SB}$ is in fact
$\rho_{SB}=\sum_{k}|v_{k}\rangle\langle v_{k}|\otimes \mathrm{Tr}_A((|v_k\rangle\langle v_k|\otimes I)\rho_{AB})$.
As the relative entropy $D(\rho\|\sigma)=\mathrm{Tr}(\rho\log\rho)-\mathrm{Tr}(\rho\log\sigma)$ is monotonic under a quantum channel
it follows that
\begin{align}\label{e:6}
&H(S|B)-H(A|B)\notag\\
=&D(\rho_{AB}\|\sum\limits_{k}(|v_k\rangle\langle v_k|\otimes I)|\rho_{AB}(|v_k\rangle\langle v_k|\otimes I))\notag\\
\geqslant&D(\rho_{RB}\|\sum\limits_{j,k}c_{jk}|u_{j}\rangle\langle u_{j}|\otimes \mathrm{Tr}_{A}((|v_k\rangle\langle v_k|\otimes I)\rho_{AB}))\notag\\
\geqslant&D(\rho_{RB}\|\sum\limits_{j}\max\limits_{k}c_{jk}|u_{j}\rangle\langle u_{j}|\otimes\rho_{B})\notag\\
=&-H(R|B)-\sum\limits_{j}p_{j}\log\max\limits_{k}c_{jk},
\end{align}
where the first expression of Eq. (\ref{e:6}) is a basic identity of the quantum relative entropy (cf. \cite{Yu, Cole}).
So the state-dependent bound under a quantum memory follows:
\begin{align}\label{e:Piani1}
H(R|B)+H(S|B)\geqslant H(A|B)-\sum\limits_{j}p_{j}\log\max\limits_{k}c_{jk}.
\end{align}
Interchanging $R$ and $S$ we also have
\begin{align}\label{e:Piani2}
H(R|B)+H(S|B)\geqslant H(A|B)-\sum\limits_{k}q_{k}\log\max\limits_{j}c_{jk}.
\end{align}

We arrange the numbers $\max\limits_{k}c_{jk}$, $j=1, \ldots, d$, in descending order:
\begin{align}\label{e:max-order}
\max\limits_{k}c_{j_1k}\geqslant \max\limits_{k}c_{j_2k}\geqslant\cdots\geqslant \max\limits_{k}c_{j_dk},
\end{align}
where $j_1j_2\cdots j_d$ is a permutation of $12\cdots d$.
Clearly $c_1=\max\limits_{k}c_{j_1k}$ and in general
$c_i\geqslant \max\limits_{k}c_{j_ik}$
for all $i$. Therefore
\begin{align}\label{e:ineq1}
&-\sum\limits_{j=1}^dp_{j}\log\max\limits_{k}c_{jk}=-\sum\limits_{i=1}^dp_{j_i}\log\max\limits_{k}c_{j_ik}\notag\\
\geqslant&-p_{j_{1}}\log c_{1}-p_{j_{2}}\log c_{2}
-\cdots-p_{j_{d}}\log c_{d}\notag\\
=&-(1-p_{j_{2}}-\cdots-p_{j_{d}})\log c_1
-p_{j_{2}}\log c_{2}
-\cdots-p_{j_{d}}\log c_{d}\notag\\
=&-\log c_{1}+p_{j_{2}}\log\frac{c_{1}}{c_{2}}+\cdots+p_{j_{d}}\log\frac{c_{1}}{c_{d}}.
\end{align}
Similarly we also have
\begin{align}\label{e:ineq2}
-\sum\limits_{k}q_{k}\log\max\limits_{j}c_{jk}
\geqslant-\log c_{1}+q_{k_{2}}\log\frac{c_{1}}{c_{2}}+\cdots+q_{k_{d}}\log\frac{c_{1}}{c_{d}},
\end{align}
for some permutation $k_{1}k_{2} \cdots k_{d}$ of $12\cdots d$.
Taking the average of Eq. (\ref{e:Piani1}) and Eq. (\ref{e:Piani2}) and plugging in Eqs. (\ref{e:ineq1})-(\ref{e:ineq2}) we have that
\begin{align}\label{e:formula}
H(R|B)+H(S|B)
\geqslant H(A|B)+\log \frac1{c_{1}}+\frac{p_{j_{2}}+q_{k_{2}}}{2}\log\frac{c_{1}}{c_{2}}
+\cdots+\frac{p_{j_{d}}+q_{k_{d}}}{2}\log\frac{c_{1}}{c_{d}}.
\end{align}
Using $p_{j_2}+q_{k_2}=\sum_{i=2}^d(p_{j_i}+q_{k_i})-\sum_{i=3}^d(p_{j_i}+q_{k_i})$ we see that Eq. (\ref{e:formula})
can be written equivalently as
\begin{align}\label{e:c2}
H(R|B)+H(S|B)
\geqslant H(A|B)+\log \frac1{c_{1}}+\frac12\sum_{i=2}^d(p_{j_i}+q_{k_i})\log\frac{c_{1}}{c_{2}}
+\frac{p_{j_{3}}+q_{k_{3}}}{2}\log\frac{c_{2}}{c_{3}}
+\cdots+\frac{p_{j_{d}}+q_{k_{d}}}{2}\log\frac{c_{2}}{c_{d}}.
\end{align}
The above transformation from Eq.(\ref{e:formula}) to Eq.(\ref{e:c2})
adds all later coefficients of $\log\frac{c_1}{c_3}, \ldots, \log\frac{c_1}{c_d}$
into that of $\log\frac{c_1}{c_2}$ and modify the argument of each $\log$ to $\log\frac{c_2}{c_3}, \ldots, \log\frac{c_2}{c_d}$.
Continuing in this way, we can write Eq.(\ref{e:c2}) equivalently as
\begin{align}\label{e:c3}
&H(R|B)+H(S|B)\notag\\
=&H(A|B)-\log c_{1}+\frac{2-(p_{j_{1}}+q_{k_{1}})}{2}\log\frac{c_{1}}{c_{2}}
+\frac{2-(p_{j_{1}}+q_{k_{1}}+p_{j_{2}}+q_{k_{2}})}{2}\log\frac{c_{2}}{c_{3}}
+\cdots+\frac{2-\sum_{i=1}^{d-1}(p_{j_{i}}+q_{k_{i}})}{2}\log\frac{c_{d-1}}{c_{d}}.
\end{align}
Since $P\oplus Q\prec\{1\}\oplus W$, we have $p_{j_{1}}+q_{k_{1}}\leqslant\Omega_{2}$, $\ldots$,
$p_{j_{1}}+q_{k_{1}}+\cdots+p_{j_{d-1}}+q_{k_{d-1}}\leqslant\Omega_{2(d-1)}$. Plugging these into Eq.(\ref{e:formula})
completes the proof. $\blacksquare$

We remark that our newly constructed bound is stronger than Coles-Piani's bound in all cases expect when the observables are mutually
unbiased (i.e. $c_{jk}=|\langle u_{j}|v_{k}\rangle|^{2}=1/d$ for any $j$, $k$).

As an example, consider the following $2\times 4$ bipartite state,  %$\rho_{AB}$
\begin{align}\label{e:state1}
\rho_{AB}=\frac{1}{1+7p}
&\left(
\begin{array}{cccccccc}
  p & 0 & 0 & 0 & 0 & p & 0 & 0 \\
  0 & p & 0 & 0 & 0 & 0 & p & 0 \\
  0 & 0 & p & 0 & 0 & 0 & 0 & p \\
  0 & 0 & 0 & p & 0 & 0 & 0 & 0 \\
  0 & 0 & 0 & 0 & \frac{1+p}{2} & 0 & 0 & \frac{\sqrt{1-p^2}}{2} \\
  p & 0 & 0 & 0 & 0 & p & 0 & 0 \\
  0 & p & 0 & 0 & 0 & 0 & p & 0 \\
  0 & 0 & p & 0 & \frac{\sqrt{1-p^2}}{2} & 0 & 0 & \frac{1+p}{2}
\end{array}
\right),
\end{align}
which is known to be entangled for $0<p<1$ \cite{Bennett}.
We take system $A$ as the quantum memory, consider the following two projective measurements:
$\{|v_k\rangle\}$ are the standard orthonormal basis on $\mathcal{H}_B$ and $\{|u_j\rangle\}$ are given by
\begin{align*}
|u_1\rangle&=(\frac{12}{\sqrt{205}}, \frac{6}{\sqrt{205}}, \frac{4}{\sqrt{205}}, \frac{3}{\sqrt{205}})^T,\\
|u_2\rangle&=(-\frac{66}{29\sqrt{205}}, \frac{172}{29\sqrt{205}}, \frac{183}{29\sqrt{205}}, -\frac{324}{29\sqrt{205}})^T,\\
|u_3\rangle&=(-\frac{11}{29\sqrt{298}}, \frac{309}{29\sqrt{298}}, -\frac{195\sqrt{\frac{2}{149}}}{29}, -\frac{27\sqrt{\frac{2}{149}}}{29})^T,\\
|u_4\rangle&=(\frac{9}{\sqrt{298}}, -\frac{9}{\sqrt{298}}, -3\sqrt{\frac{2}{149}}, -5\sqrt{\frac{2}{149}})^T.
\end{align*}
Then the overlap matrix $(|\langle u_{j}|v_{k}\rangle|^{2})_{jk}$ is given by
\begin{equation}\label{e:overlap}
\left(
\begin{array}{cccc}
  \frac{144}{205} & \frac{36}{205} & \frac{16}{205} & \frac{9}{205} \\
  \frac{4356}{172405} & \frac{29584}{172405} & \frac{33489}{172405} & \frac{104976}{172405} \\
  \frac{121}{250618} & \frac{95481}{250618} & \frac{76050}{125309} & \frac{1458}{125309} \\
  \frac{81}{298} & \frac{81}{298} & \frac{18}{149} & \frac{50}{149}
\end{array}
\right).
\end{equation}
Thus $\Omega_{4}\neq2$ and $c_{2}\neq c_{3}$.
The comparison between Coles-Piani's bound and Eq. (\ref{e:result2}) in the presence of quantum memory is displayed in FIG. 1,
which shows that our new bound is strictly tighter for all $p\in (0, 1)$.

\begin{figure}
\centering
\includegraphics[width=0.45\textwidth]{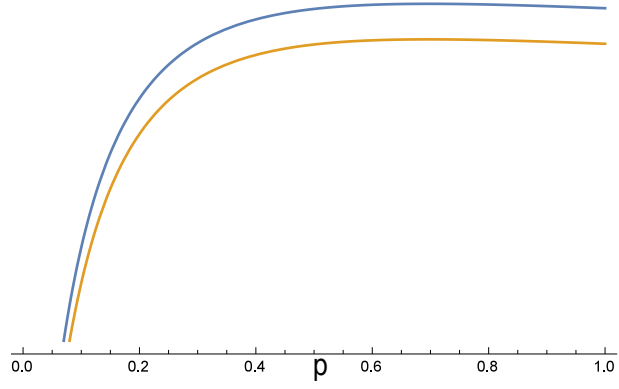}
\caption{Comparison of bounds for entangled $\rho_{AB}$. The upper blue curve is the new bound \eqref{e:result2}
and the lower yellow curve is Coles-Piani's bound.}
\end{figure}

After presenting the general result, we now turn to its special situation for the state-independent
bound in the absence of quantum memory. As many state-independent bounds cannot be generalized to
the general situation, a separate treatment is needed and one will see that our new
bound fares reasonably well even in the absence of quantum memory.

\noindent\textbf{Corollary.} {\it Let
$R=\{|u_{j}\rangle\}$ and $S=\{|v_{k}\rangle\}$ be any two orthonormal bases on $d$-dimensional
Hilbert space $\mathcal{H}_{A}$. Then for any state $\rho_{A}$ over $\mathcal{H}_{A}$, we have the following
inequality:
\begin{align}\label{e:result1}
H(R)+H(S)
\geqslant\log \frac1{c_{1}}+\frac{1-\sqrt{c_{1}}}{2}\log\frac{c_{1}}{c_{2}}+\frac{2-\Omega_{4}}{2}\log\frac{c_{2}}{c_{3}}
+\frac{2-\Omega_{6}}{2}\log\frac{c_{3}}{c_{4}}+\cdots+\frac{2-\Omega_{2(d-1)}}{2}\log\frac{c_{d-1}}{c_{d}},
\end{align}
where $\Omega_{k}$ and $c_{i}$ are defined in Eq.(\ref{e:result2}).}

The corollary can be similarly proved due to the following simple observation.
When measurements are performed on system $A$,
$H(R)+H(S)\geqslant-\sum\limits_{j}p_{j}\log\sum\limits_{k}q_{k}c_{jk}+H(A)
\geqslant-\sum\limits_{j}p_{j}\log\max\limits_{k}c_{jk}+H(A)$.
Then the corollary follows directly from the proof of the theorem.

We now compare our bound given in the corollary with
some of the well-known bounds in this special case. First of all,
 our new bound is clearly tighter than Coles-Piani's bound $B_{CP}=\log \frac1{c_{1}}+\frac{1-\sqrt{c_{1}}}{2}\log\frac{c_{1}}{c_{2}}$ as we
 have already shown mathematically.
In \cite{Rudnicki} Rudnicki {\it et al.} obtained the {\it direct-sum majorization} bound, one of the major ones for the state-independent states, thus we will focus on comparing our bound with this one.
Consider the following unitary matrices between two measurements
\begin{align}\label{e:u}
U(\theta)=M(\theta)OM(\theta)^{\dag},
\end{align}
where $\theta\in [0, \pi/4]$ and
\begin{align}\label{e:theta}
M(\theta)=
&\left(
\begin{array}{ccc}
  1 & 0 & 0 \\
  0 & \cos\theta_{1} & \sin\theta_{1} \\
  0 & -\sin\theta_{1} & \cos\theta_{1}
\end{array}
\right),
\end{align}
with the overlap matrix
\begin{align}
O=
&\left(
\begin{array}{ccc}
  0.4575 & 0.4575 & 0.7625 \\
  -0.2453 & 0.8892 & -0.3863 \\
  -0.8547 & -0.0103 & 0.5190
\end{array}
\right).
\end{align}

\begin{figure}
\centering
\includegraphics[width=0.45\textwidth]{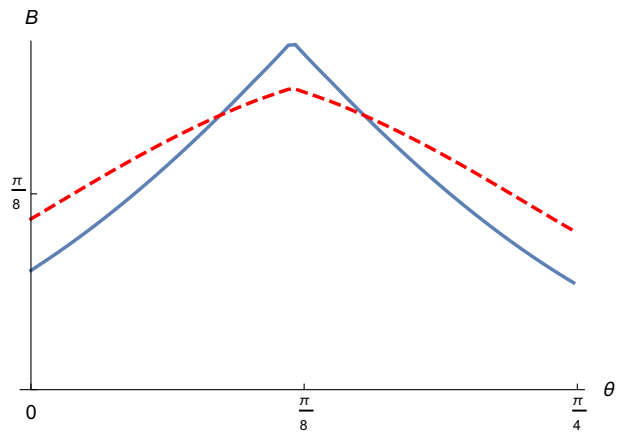}
\caption{Comparison of direct-sum majorization bound (dashed red) and Eq. (\ref{e:result1}) for unitary matrices $U(\theta)$.}
\end{figure}

FIG. 2 shows that the direct-sum majorization bound and our bound Eq. (\ref{e:result1}) are complementary, as there are regions
where our bound is tighter.
This implies that our bound in this
special case is also nontrivial. We would like to remark that there are other bounds for the entropic uncertainty relations in the absence of quantum side
information, but most of them cannot be generalized to the case with quantum memory.

Note that the overlaps, which are commonly used in entropic uncertainty relations, are state-independent measurements so we can only
consider them when the experimental device is trusted. For {\it device-independent uncertainty} based on state-dependent and
incompatible measures, see \cite{Kaniewski}. There seems no clear relation between our bound and the bound based on state-dependent anticommutators, and it is still open whether the approach of \cite{Kaniewski} based on state-dependent measurements can be extended to
allow for the quantum side information. On the other hand, our result holds for the general case with the quantum side information.

{\bf Conclusion.} We have found new lower bounds for the sum of the entropic uncertainties in the presence of quantum memory. Our new bounds have formulated the complete dependence on all $d$ largest entries in the overlap matrix between two measurements
on a $d$-dimensional Hilbert space, while the previously best-known
bound involves with the first two largest entries.
We have shown that our bounds are strictly tighter than previously known entropic uncertainty bounds
with quantum side information by mathematical argument in the general situation. In the special case without quantum memory, our bound also
offers significant new information as it is complementary to some of the best known bounds in this situation. Moreover, as entropic uncertainty relations in the presence of quantum memory have a wide range of applications, our results are expected to shed new lights on investigation of quantum information processing such as information exclusive relation \cite{Xiao}, entanglement detection, quantum key distribution and other cryptographic scenarios.

\medskip
\noindent{\bf Acknowledgments}\, \, The work is supported by
NSFC (grant Nos. 11271138, 11531004, 11275131¢G?11675113), CSC and Simons Foundation grant 198129.


\bigskip
\begin{thebibliography}{9}

\bibitem{Heisenberg} W. Heisenberg,
Z. Phys. \textbf{43}, 172 (1927).

\bibitem{Robertson} H. P. Robertson,
Phys. Rev. \textbf{34}, 163 (1929).

\bibitem{Birula} I. Bia{\l}ynicki-Birula and L. Rudnicki,
Statistical Complexity, ed. by K. Sen,
Springer Netherlands, Dordrecht  pp. 1-34 (2011).

\bibitem{Marco} P. J. Coles, M. Berta, M. Tomamichel, and S. Wehner,
arXiv: 1511.04857v1.

\bibitem{BBM} I. Bialynicki-Birula and J. Mycielski, Comm. Math. Phys. 44, 129 (1975).

\bibitem{Deutsch} D. Deutsch,
Phys. Rev. Lett. \textbf{50}, 631 (1983).

\bibitem{Kraus} K. Kraus,
Phys. Rev. D. \textbf{35}, 3070 (1987).

\bibitem{Maassen} H. Maassen and J. B. M. Uffink,
Phys. Rev. Lett. \textbf{60}, 1103 (1988).

\bibitem{B} M. Berta, M. Christandl, R. Colbeck, J.~M. Renes, and R. Renner,
Nat. Phys. \textbf{6}, 659 (2010).

\bibitem{Coles} P. J. Coles and M. Piani,
Phys. Rev. A \textbf{89}, 022112 (2014).

\bibitem{Rudnicki} {\L}. Rudnicki, Z. Pucha{\l}a, and K. \.{Z}yczkowski,
Phys. Rev. A \textbf{89}, 052115 (2014).

\bibitem{Puchala} Z. Pucha{\l}a, {\L}. Rudnicki, and K. \.{Z}yczkowski,
J. Phys. A \textbf{46}, 272002 (2013).

\bibitem{Yu} P. J. Coles, L. Yu, V. Gheorghiu, and R. B. Griffiths,
Phys. Rev. A \textbf{83}, 062338 (2011).

\bibitem{Cole} P. J. Coles,
Phys. Rev. A \textbf{85}, 042103 (2012).

\bibitem{Bennett} C. H. Bennett, D. P. DiVincenzo, T. Mor, P. W. Shor, J. A. Smolin, and B. M. Terhal,
Phys. Rev. Lett. \textbf{82}, 5385 (1999).

\bibitem{Kaniewski} J. Kaniewski, M. Tomamichel, and S. Wehner,
Phys. Rev. A \textbf{90}, 012332 (2014).

\bibitem{Xiao} Y. Xiao, N. Jing, and X. Li-Jost,
Sci. Rep. \textbf{6}, 30440 (2016).

%\bibitem{Ghirardi} G. Ghirardi, L. Marinatto, and R. Romano,
%Phys. Lett. A \textbf{317}, 32 (2003).

\end{thebibliography}
\end{document}